\documentclass[iop,twocolumn,apj]{emulateapj}
\usepackage{aas_macros}

\newcommand{\abs}[1]{{\left\vert{#1}\right\vert}}
\newcommand{\vesc}{v_{\mathrm{esc}}}
\newcommand{\vmin}{v_{\mathrm{min}}}

\shorttitle{VDF of Dark Matter from Simulations}
\shortauthors{Mao et al.}

\begin{document}

\title{Halo-to-Halo Similarity and Scatter in the Velocity Distribution of Dark Matter}
\author{Yao-Yuan Mao, Louis E.~Strigari, Risa H.~Wechsler, Hao-Yi Wu\altaffilmark{1}, Oliver Hahn} 
\affil{Kavli Institute for Particle Astrophysics and Cosmology
  \& Physics Department, Stanford University, Stanford, CA 94305}
\affil{SLAC National Accelerator Laboratory, Menlo Park, CA, 94025}

\altaffiltext{1}{Currently at Physics Department, The University of Michigan, Ann Arbor, MI 48109}

\begin{abstract}
We examine the velocity distribution function~(VDF) in dark matter halos
from Milky Way to cluster mass scales. We identify an empirical model for
the VDF with a wider peak and a steeper tail than a Maxwell--Boltzmann
distribution, and discuss physical explanations. We quantify sources of scatter
in the VDF of cosmological halos and their implication for direct detection of 
dark matter. Given modern simulations and observations, we find that the most
significant uncertainty in the VDF of the Milky Way arises from the unknown radial
position of the solar system relative to the dark matter halo scale radius.
\end{abstract} 


\maketitle

\section{Introduction}
Dark matter is the dominant component of matter in the Universe, and the key to the formation of large-scale and galactic structures.  Modern cosmological observations suggest that dark matter is composed of a yet-unidentified elementary particle~(e.g.~\citealt{2010ARA&A..48..495F}).  However, direct evidence for dark matter particles has proved elusive.  Experiments that search for Weakly Interacting Massive Particles (WIMPs), one of the most plausible particle dark matter candidates, seek to identify the scattering of a WIMP with a nucleus in an underground detector~\citep{2008EPJC...56..333B,2010Sci...327.1619C,2011PhRvL.106m1301A,2012EPJC...72.1971A,2011PhRvL.107m1302A}.  Constraining, and eventually measuring, the WIMP mass and cross section requires a precise understanding of the dark matter spatial and velocity distribution at the Earth's location in the Milky Way~\citep{2009JCAP...11..019S,2010PhRvD..82b3530M,2011MNRAS.415.3177R,2012MPLA...2730004G}. 

Dark matter is distributed in halos extending beyond the visible components of galaxies; many statistical properties including the formation and structure of these halos have been well characterized by simulations.   Despite the diversity in the merger and accretion histories of dark matter halos of different masses, cosmological simulations have long suggested near universality in the density profiles of halos~\citep{1996ApJ...462..563N,1997ApJ...490..493N}.  There have been several attempts to connect this universality in the density profile to the dark matter Velocity Distribution Function~(VDF)~\citep{2006JCAP...01..014H,2010JCAP...02..030K,2010MNRAS.402...21N}.  However, there is no well-established model or description for the VDF that has been rigorously tested with cosmological simulations.

Both the implications for direct detection and the quest for a theoretical understanding of the phase-space distribution in dark matter halos motivate a study of the VDF.  Under specific, and perhaps too stringent, assumptions, including isolation, equilibrium, spherical symmetry, and isotropy, the VDF may be determined uniquely from the density profile.  For example, with all assumptions named above and a known density profile, the ergodic distribution function can be calculated using Eddington's formula~\citep{1916MNRAS..76..572E}.  Although useful as an analytic framework, these assumptions are unlikely to strictly hold for halos formed via hierarchical merging.

In absence of an understanding from first principles, a practical approach to study the VDF involves appealing directly to dark matter halos with a wide range of physical properties in cosmological simulations.  Quantifying the VDF directly from cosmological simulations would provide a better empirically-motivated framework to predict signals in direct detection experiments.  Furthermore, with a parametrized VDF, it becomes more tractable to study the relations between the VDF and other physical quantities of the halos, such as mass, density profile, shape, and formation history.

In this study, we use a suite of dark matter halos from cosmological simulations to study the VDFs at different radii of these halos. We identify a similarity in VDFs among a wide range of halos with different masses, concentrations, and other physical quantities, that depends primarily on $r/r_s$, the radius at which it is measured divided by the scale radius of the density profile.  We further notice that neither standard Maxwell--Boltzmann models~\citep{1996APh.....6...87L} nor models that have been previously proposed to describe collisionless structures~\citep{2006JCAP...01..014H,2010JCAP...02..030K,2010MNRAS.402...21N} are able to provide an adequate description of cosmological VDFs.  Instead, we describe the distribution of the norm of velocity (in the Galactic rest frame) more accurately with an empirical model: 
\begin{equation}
  \label{eq:vdf} 
    f(\abs{\mathbf{v}})=\left\{\begin{array}{ll}
	    A\exp(-\abs{\mathbf{v}}/v_0)\left(\vesc^2-\abs{\mathbf{v}}^2\right)^p,&0\leq\abs{\mathbf{v}}\leq \vesc\\
		0,&\mathrm{otherwise}, 
	\end{array}\right. 
\end{equation}
where the normalization constant~$A$ is chosen such that the integral $4\pi\int_0^{\vesc}v^2f(v)dv$ equals the number of particles in the region of interest.  Note that in this parameterization the VDF approaches an exponential distribution instead of a Gaussian distribution at the low-velocity end.  With this model, we quantify the scatter in the VDF from a variety of sources, including halo-to-halo scatter, scatter from finite particle sampling, and scatter from the uncertain position of the Earth within a given halo.  We further identify the largest uncertainties that currently exist in our understanding of the VDF at the location of the Earth in our Galaxy, and quantify their relevance for inferences from direct detection experiments.

\section{Universal Velocity Distribution in Simulations} 
To identify the relevant physical quantities which affect the VDF and to quantify scatter in the distributions among different halos in cosmological simulations, we must examine a large number of halos across a wide range of mass.  We also need high resolution to reduce sampling error and distinguish differences in VDFs for different parameters.

In this study, we use halos from the {\sc Rhapsody} and {\sc Bolshoi} simulations; state-of-the-art dark-matter-only simulations with high mass resolution.  {\sc Rhapsody} consists of re-simulations of 96 massive cluster-size halos with $M_{\mathrm{vir}}=10^{14.8\pm 0.05}M_\odot h^{-1}$.  The particle mass is $1.3 \times 10^8M_\odot h^{-1}$, resulting in $\sim 5\times 10^6$ particles in each halo.  This simulation set currently comprises the largest number of halos simulated with this many particles in a narrow mass bin (Fig.~1 of~\citealt{2012arXiv1209.3309W}).  {\sc Bolshoi} is a full cosmological simulation, with similar mass resolution, $ 1.3\times 10^8M_\odot h^{-1}$.  For detailed descriptions of the {\sc Rhapsody} and {\sc Bolshoi} simulations, refer to~\citet{2012arXiv1209.3309W} and~\citet{2011ApJ...740..102K} respectively.

We use the phase-space halo finder {\sc Rockstar}~\citep{2011arXiv1110.4372B} to identify host halos at $z=0$.  The masses and radii of the halos are defined by the spherical overdensity of virialization, $M(<r_{\mathrm{vir}}) = \frac{4\pi}{3} r_{\mathrm{vir}}^3 \Delta_{\mathrm{vir}} \rho_c$, where $\Delta_{\mathrm{vir}}=94$ and $\rho_c$ is the critical density.  We examine the VDFs at a range of radii.  A VDF at radius $r$ uses all particles within a spherical shell centered at the halo center with the inner and outer radii of $10^{\pm 0.05}r$, so that the ratio of the shell width to the radius is fixed.  In each shell, we assign the escape velocity ($\vesc$) as the spherically-averaged $\vesc$ of all particles in the shell.  We have verified that $\vesc$ determined from this method is consistent with the same quantity deduced from the best-fitting spherically-averaged smooth density profile.

We fit each halo with an NFW density profile, 
\begin{equation}
  \label{eq:nfw} 
	\rho(r)=\frac{\rho_s}{(r/r_s)(1+r/r_s)^{2}}, 
\end{equation}
where $r_s$ is the scale radius at which the log--log slope is $-2$.  The fit uses maximum-likelihood estimation based on particles within $r_\mathrm{vir}$.  The halo concentration is defined as $c=r_\mathrm{vir}/r_s$.

\begin{figure}
\centering \includegraphics{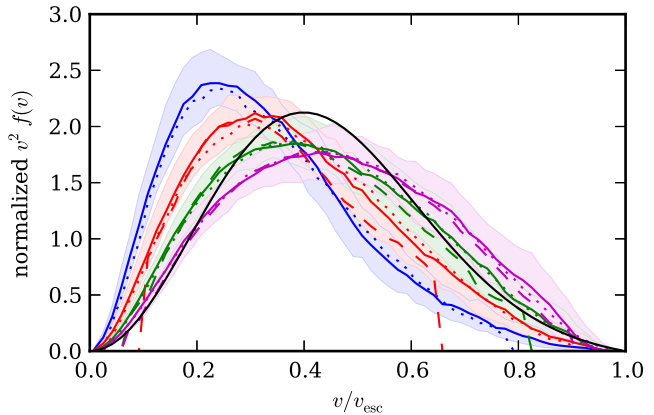}
\caption{Solid colored lines show the stacked velocity distribution for 96 halos in \textsc{Rhapsody}, at different values of $r/r_s$: (from left to right) 0.15 (blue), 0.3 (red), 0.6 (green), 1.2 (magenta).  Bands show the 68\% halo-to-halo scatter in those VDFs.  Dashed and dotted colored lines indicate the same values of $r/r_s$ in \textsc{Bolshoi} with halos of $M_{\mathrm{vir}}\sim 10^{12}$ and $10^{13}M_\odot h^{-1}$ respectively. The VDFs of low-mass halos are cut at the head and tail due to limited particle number, and their scatter is not shown.  The SHM ($v_0=220$ km/s and $\vesc =544$ km/s) is shown for comparison (black).}
  \label{fig:vdf_stack_radius}
\end{figure}

\placefigure{fig:vdf_stack_radius}

Fig.~\ref{fig:vdf_stack_radius} shows the VDF at different values of $r/r_s$.  The value of $r/r_s$ affects the shape of VDF dramatically.  The peak of the distribution is a strong function of $r/r_s$.  If instead the velocity is normalized by the circular velocity at each radius rather than the escape velocity, this trend will be slightly weakened but still significant.  This trend in $r/r_s$ is not surprising because the VDF heavily depends on the gravitational potential.  If the density profiles of simulated halos can be described by the NFW profile, which is a function of $r/r_s$ only (up to a normalization constant), the VDF should mostly depend on $r/r_s$ until the isolated NFW potential breaks down at large radius.

The above trend is robust for halo masses down to $\sim 10^{12}$ M$_\odot$, as shown by the \textsc{Bolshoi} simulation in Fig.~\ref{fig:vdf_stack_radius}.  The scatter of the VDFs in the low-mass halos considered is somewhat larger due to resolution.  However, when the high-mass halos are downsampled to have the same particle number, the spreads in the stacked VDF are comparable to the low-mass halos.  We further investigated the impact of a variety of parameters characterizing the halo on the shape of the VDF, and found that for a fixed value of $r/r_s$, the halo-to-halo scatter in the VDFs is not significantly reduced when binning on concentration, shape, or formation history.  A detailed discussion on this halo-to-halo scatter is in Section \ref{sec:scatter}.

\section{Models of the Velocity Distribution Function}
The dark matter velocity distribution in halos is set by a sequence of mergers and accretion.  The process of violent relaxation~\citep{1967MNRAS.136..101L} may be responsible for the resulting near-equilibrium distributions observed in dark matter halos and in galaxies.  These near-equilibrium distributions explain why existing VDF models (see e.g.~\citealt{2012JCAP...01..024F}), including the Standard Halo Model (SHM), King model, the double power-law model, and the Tsallis model, are all variants of the Maxwell--Boltzmann distribution.  Recent studies have shown that the widely-used SHM, which is a Maxwell--Boltzmann distribution with a cut-off put in by hand, is inconsistent with the VDF found in a handful of individual simulations~\citep{2003PhRvL..90u1301S,2009MNRAS.395..797V, 2010JCAP...02..030K,2012JCAP...08..027P} and in the study of rotation curve data \citep{2012arXiv1210.2328B}.  The double power-law model was proposed to suppress the tail of the distribution, by raising the SHM to the power of a parameter $k$~\citep{2011PhRvD..83b3519L}.  The Tsallis model replaces the Gaussian in Maxwell--Boltzmann distribution with a $q$-Gaussian, which approaches to a Gaussian as $q\rightarrow 1$~\citep{2008PhRvD..77b3509V}.  It was argued that the Tsallis model provides better fit to simulations with baryons~\citep{2010JCAP...02..012L}, although this conclusion may be affected by the relatively low resolution of the simulations. 

\begin{figure}
\centering \includegraphics{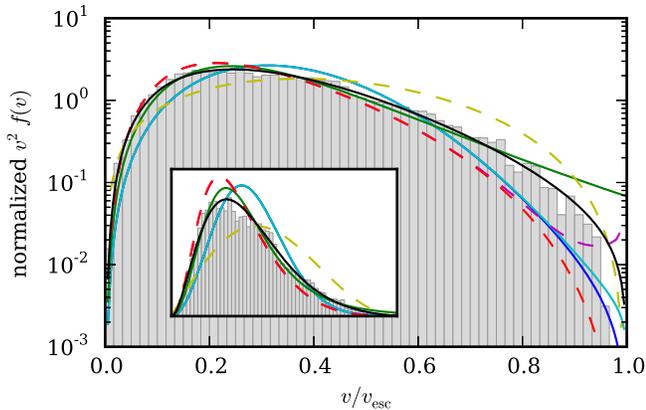} 
\caption{The VDF for one representative dark matter halo in {\sc Rhapsody} (histogram), along with the best fits using Eq.~(\ref{eq:vdf}) with $(v_0/\vesc,p)=(0.13,0.78)$ (black, $\chi^2=0.59$), SHM (blue, 9.67), the double power-law model (cyan, 9.47), the Tsallis model (green, 1.99), and the analytic VDFs from Eddington's formula with isotropic assumption (red dash, 8.48), Osipkov--Merritt (magenta dash, 6.41), and constant $\beta=1/2$ (yellow dash, 11.8).  The $y$-axis is in log scale in the main figure and linear in the inset.} 
  \label{fig:vdf_compare} 
\end{figure}

In contrast, our empirical model, Eq.~(\ref{eq:vdf}), is not based on a Gaussian distribution but rather on an exponential distribution.  It also has a power-law cut-off in (binding) energy.  Fig.~\ref{fig:vdf_compare} shows the VDF in a simulated halo, along with the best fit from Eq.~(\ref{eq:vdf}) and the best fits from other conventional models.  All the best-fit parameters are obtained from the maximum-likelihood estimation in the range of $(0, \vesc)$.  The fits using Eq.~(\ref{eq:vdf}) are statistically better than other models or the analytic VDFs, especially around the peak and the tail.  We performed the likelihood-ratio test and found that our model fits significantly better for {\it all} {\sc Rhapsody} halos than the SHM or the double power-law model at all four radii shown in Fig.~\ref{fig:vdf_stack_radius}. 

In Fig.~\ref{fig:vdf_compare} we also compare three analytic VDFs.  For the isotropic model shown, the analytic VDF is given by Eddington's formula, which gives a one-to-one correspondence between the density profile and the VDF.  For anisotropic systems, one must also model the anisotropy parameter, defined as $\beta=1-(\sigma_\theta^2+\sigma_\phi^2)/(2\sigma_r^2)$, where $\sigma^2$ is the variance in each velocity component.  There is currently no analytic VDF whose anisotropy profile matches that measured in simulations, so we choose three simple and representative anisotropic models: constant anisotropy (with $\beta=0$ and $1/2$) and the Osipkov--Merritt model~\citep{1979PAZh....5...77O,1985AJ.....90.1027M}.  The phase-space distributions of these models can be determined numerically~\citep{2008gady.book.....B}.  For all three cases, we adopt the NFW profile as in Eq.~(\ref{eq:nfw}), with the best-fit scale radius.  For the Osipkov--Merritt model, we use the best-fit anisotropy radius.  It is shown in Fig.~\ref{fig:vdf_compare} and also suggested by the chi-square test for the models considered that the analytic VDFs do not describe the simulated VDF well.

\placefigure{fig:vdf_compare}

\begin{figure}
\centering \includegraphics{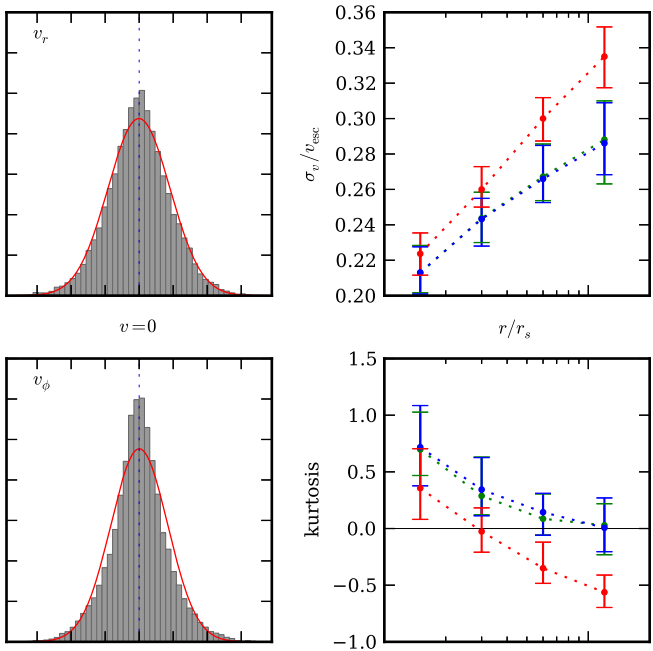} 
\caption{\textit{Left}: The histograms of $v_r$ and $v_\phi$ of the same halo shown in Fig.~\ref{fig:vdf_compare}, with the best-fit normal distributions (red lines).  \textit{Right}: The velocity dispersion $\sigma_v/\vesc$ and the kurtosis, along the three axes: $v_r$ (red), $v_\theta$ (green), and $v_\phi$ (blue). Both the dispersion and the kurtosis are measured in spherical shells at different $r/r_s$ and averaged over all halos in {\sc Rhapsody}, with the error bars showing the 68\% halo-to-halo scatter. The dashed lines are only to guide the eyes.} 
  \label{fig:vdf_gaussian} 
\end{figure}

Our VDF model, Eq.~(\ref{eq:vdf}), consists of two terms: the exponential term and the cut-off term.  The origin of the the exponential term can be explained by the anisotropy in velocity space.  Fig.~\ref{fig:vdf_gaussian} shows the distributions, the dispersion, and the kurtosis of the velocity vectors along the three axes of the spherical coordinate.  Kurtosis is a measure of the peakedness of a distribution, defined as $(\sum_i v_i^4)/(\sum_i v_i^2)^2 - 3 $, where $v_i$ is the velocity of the $i$-th particle along one axis, and this value is zero for the normal distribution.  The ratios of dispersion between the three axes are close to one at small radii, and the ratios increase with radius.  The kurtosis, on the other hand, is in general non-zero and decreases with radius.  An important consequence of the non-zero kurtosis is that even if the dispersion along the three axes are similar (anisotropy parameter $\beta\sim 0$), the velocity vectors do \textit{not} follow an isotropic multivariate normal distribution in any coordinate system (even after a local coordinate transformations).  In other words, as long as there exists either anisotropy or non-zero kurtosis in a certain coordinate, the norms of the velocity vectors will not follow the Maxwell--Boltzmann distribution.  Indeed, Fig.~\ref{fig:vdf_gaussian} shows that in the simulations, one always has non-zero kurtosis and/or anisotropy.  Other simulations also indicate that the velocity vectors of dark matter particles have anisotropy~\citep{2011arXiv1111.3944A,2012JCAP...10..049S} and non-zero kurtosis~\citep{2009MNRAS.395..797V}.  We further found that if the ratios of dispersion between the three axes of a multivariate normal distribution are around 0.2 to 0.6, the norms of those random vectors will follow a distribution which resembles our model without the cut-off term, $v^2 \exp(-v/v_0)$. (For a formal discussion on this topic, see e.g.~\citealt{2009ITSP...57.4027B}.)  This suggests that if one can find a coordinate system where the distributions of the velocity components are all distributed normally (with zero kurtosis), there will be a larger difference between the dispersion along the three axes in this new coordinate system than in the spherical coordinate.

\placefigure{fig:vdf_gaussian}

The $(\vesc^2-v^2)^p$ term in our VDF model introduces a cut-off at the escape velocity. It further suppresses the
VDF tail more than the exponential term alone does.  Despite that this cut-off term has the form of a power-law in (binding) energy, the best-fit values of the parameter $p$ does not necessarily reflect the ``asymptotic'' power-law index $k$, defined as $k=\lim_{\mathcal{E}\rightarrow 0}(d\ln f/d\ln\mathcal{E})$, where $f(\mathcal{E})$ is the (binding) energy distribution function.  The relation between $k$ and the outer density slope has been studied in the literature~\citep{2006PhRvD..73b3524E,2011PhRvD..83b3519L}.  However, because $d\ln f/d\ln\mathcal{E}$ deviates from its asymptotic value $k$ rapidly as $\mathcal{E}$ deviates from zero, the asymptotic power-law index $k$ could be very different from the best-fit power-law index for the VDF tail (e.g.~$v>0.9\vesc$).  Furthermore, the shape of the VDF power-law tail could be set by recently-accreted subhalos that have not been fully phase-mixed~\citep{2012PhRvD..86f3505K}, and hence has no simple relation with the density profile.  In high-resolution simulated dark matter halos, particles stripped off of a still-surviving subhalo are seen to significantly impact the tail of the VDF.  A larger sample of simulations at higher resolution than we consider in the current analysis will be needed to further test this hypothesis.

\section{Halo-to-halo Scatter in Velocity Distributions}
  \label{sec:scatter}
  
\begin{figure}
\centering \includegraphics{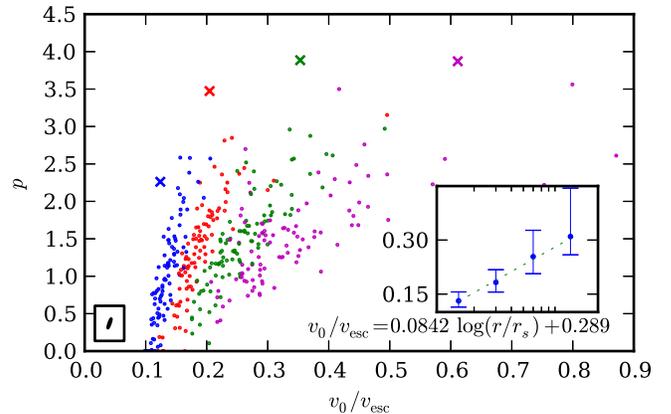} 
\caption{Distribution of the best-fit parameters, $v_0$ and $p$ of Eq.~(\ref{eq:vdf}), which describes the simulated VDFs.  Each dot represents one halo from the {\sc Rhapsody} simulation at a certain $r/r_s$: (from left to right) 0.15 (blue), 0.3 (red), 0.6 (green), 1.2 (magenta).  The cross symbols show the best-fit parameters to isotropic analytic VDFs obtained from Eddington's formula at corresponding radii.  The typical uncertainty of the fit is shown in the lower left corner.  The lower right inset shows the linear relation between $v_0/\vesc$ and $\log(r/r_s)$.}
  \label{fig:vdf_para_scatter} 
\end{figure}

We demonstrated above that there exists a similarity in VDFs for a wide range of simulated dark matter halos; Eq.~(\ref{eq:vdf}) provides a good description of this similarity.  We now quantify explicitly how the VDF depends on $r/r_s$ and the associated halo-to-halo scatter.  Fig.~\ref{fig:vdf_para_scatter} shows a scatter plot of the velocity distributions for different halos, characterized by the two parameters of Eq.~(\ref{eq:vdf}), for different $r/r_s$.  The regions of $(v_0,p)$ parameter space for different $r/r_s$ are distinct, which implies that $r/r_s$ is the most relevant quantity in determining the shape of the velocity distribution.  We also found that the parameter $v_0/\vesc$ has a linear relationship in $\log(r/r_s)$, as shown in the inset of Fig.~\ref{fig:vdf_para_scatter}.  

\placefigure{fig:vdf_para_scatter}

We note that there is significant degeneracy between the two parameters $(v_0, p)$.  This degeneracy comes from the fact that a larger value of $p$ is needed to steepen the tail of the VDFs which have larger values of $v_0$. In our fitting process we left both parameters free because there is no simple relation between $v_0$ and $p$ for all radii.  Because of this degeneracy, there also exists a linear relation between $p$ and $\log(r/r_s)$.  However, since the best-fit $p$ is not well-constrained due to the low number of particles in the tail of the VDF, the 
relation between $p$ and $\log(r/r_s)$ is not well determined either.

\begin{figure}
\centering \includegraphics{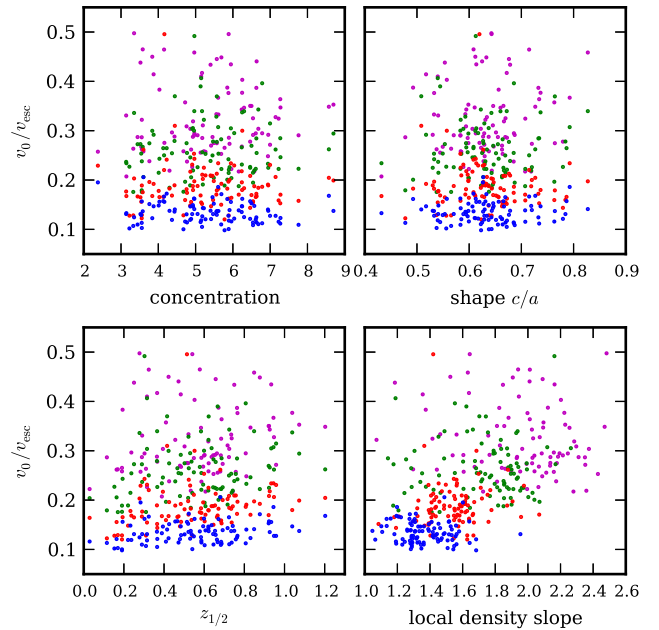} 
\caption{Scatter plots of the best-fit parameter $v_0/\vesc$ with concentration, halo shape ($c/a$), formation time ($z_{1/2}$), and local density slope on the $x$-axes respectively. Each dot represents one halo from the {\sc Rhapsody} simulation at a certain $r/r_s$: (from bottom to top in each panel) 0.15 (blue), 0.3 (red), 0.6 (green), 1.2 (magenta). For any fixed $r/r_s$, there is no significant correlation between $v_0/\vesc$ and the aforementioned quantities on the $x$-axes. See text for details.}
  \label{fig:vdf_scatter_other} 
\end{figure}

\placefigure{fig:vdf_scatter_other}

In Fig.~\ref{fig:vdf_para_scatter} we see there exists halo-to-halo scatter even for a fixed $r/r_s$.  This intrinsic scatter could arise from the statistics of the samples or some other physical quantities.  
Fig.~\ref{fig:vdf_scatter_other} shows the best-fit $v_0/\vesc$ at different radii as a function of concentration, halo shape ($c/a$), formation time ($z_{1/2}$), and local density slope ($-d\ln\rho/d\ln r$) respectively, as defined in~\citet{2012arXiv1209.3309W}.  We found that at a given $r/r_s$ (a fixed color), $v_0/\vesc$ does not have a significant correlation with the physical quantities on the $x$-axis (except for $z_{1/2}$ in the two smallest radial bins).  This reinforces the main result of this study: the VDF is mostly determined by $r/r_s$ (i.e. the gravitational potential).  We note that the lower left panel of Fig.~\ref{fig:vdf_scatter_other} shows a weak correlation between $z_{1/2}$ and $v_0/\vesc$; however if the halos with $z_{1/2} < 0.25$ are removed, this correlation is no longer statistically significant.  Halos with recent accretion tend to have larger deviations from the NFW profile, and this results in a slight overestimate of the best-fit scale radius (fit to an NFW profile).  We do not expect the Milky Way has had a recent major merger with $z_{1/2} < 0.25$.  This indicates that for possible Milky Way host halos, one can exclude these systems with recent major mergers, and there will be no remaining correlation between formation time and $v_0$.

For Milky Way size halos, it has been suggested that the VDF has a universal shape depending only on the velocity dispersion and the local density slope~\citep{2006JCAP...01..014H}.  This is related to our finding in a way that the magnitude of the velocity dispersion is roughly proportional to $\vesc$ and the local density slope for an NFW profile is given by a monotonic function of $r/r_s$, 
\begin{equation}
  \label{eq:local_slope}
	\frac{d\ln\rho}{d\ln r}=-\frac{1+3(r/r_s)}{1+(r/r_s)}.
\end{equation}
However, our study suggests that $r/r_s$ is a more fundamental quantity than the local density slope in determining the shape of the VDF.  Fig.~\ref{fig:vdf_scatter_other} illustrates that $v_0/\vesc$ does not grow with the local density slope when one only looks at a fixed $r/r_s$ (points with the same color), but it does grow with $r/r_s$ when the local density slope is fixed.

\section{Implications for Direct Detection Rates} 
Given the known dependence on $r/r_s$, we can now examine the impact on direct dark matter detection experiments.  The differential event rate per unit detector mass of dark matter interactions in direct detection experiments is 
\begin{equation}
	\left.\frac{dR}{dQ}\right\vert_{Q}=\frac{\rho_0\sigma_0}{2\mu^2m_\mathrm{dm}}A^2\left\vert F(Q)\right\vert^2\int_{\vmin(Q)}d^3v\,\frac{f(\mathbf{v}+\mathbf{v_e})}{v},
	\label{eq:dd_rate_particle} 
\end{equation}
where $Q$ is the recoil energy, $\rho_0$ is the local dark matter density, $\sigma_0$ is the WIMP-nucleus cross section at zero momentum transfer, $m_\mathrm{dm}$ is the WIMP mass, $\mu$ is the WIMP-nucleus reduced mass, $A$ is the atomic number of the nucleus, $\left\vert F(Q)\right\vert^2 $ is the nuclear form factor, $\vmin=(Q m_N/2\mu^2)^{1/2}$ for an elastic collision, $f$ is the VDF in the Galactic rest frame, and $\mathbf{v_e}$ is the velocity of Earth in the Galactic rest frame~\citep{1996APh.....6...87L}.

\begin{figure}
\centering \includegraphics{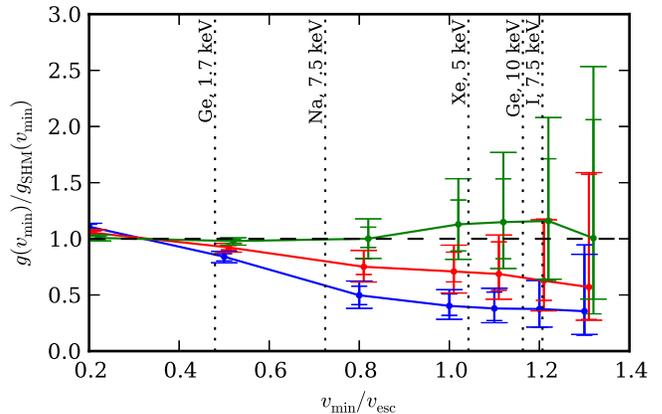} 
\caption{Ratio of detection rate predicted by Eq.~(\ref{eq:vdf}) with parameters obtained from {\sc Rhapsody}, for different $r/r_s$: (from bottom to top) 0.15 (blue), 0.3 (red), 0.6 (green), to that of the SHM with conventional parameters. Vertical dotted lines show $\vmin$ for different detectors: (from left to right) CoGeNT, DAMA-Na, XENON, CDMS, DAMA-I, expressed in (nucleus, threshold energy)~\citep{2011PhRvL.106m1301A,2008EPJC...56..333B,2011PhRvL.107m1302A,2010Sci...327.1619C}, assuming a WIMP mass of 10 GeV.  The error bars show the 68\% halo-to-halo scatter, and those with wider caps include the scatter in different directions.  The $x$-axis is slightly offset for clarity. The lines which connect the data points are only to guide the eyes.}
  \label{fig:vdf_rate_vmin} 
\end{figure}

With Eq.~(\ref{eq:dd_rate_particle}) one can calculate the event rate given VDF and $\vmin$.  We calculated this rate for each halo using the best-fit exponential model of the VDF, for different $\vmin$ and different $r/r_s$.  The results are shown in Fig.~\ref{fig:vdf_rate_vmin}, where we divided the rate by the rate calculated from the SHM with conventional parameters $v_0=220$ km/s and $\vesc=544$ km/s for comparison.

\placefigure{fig:vdf_rate_vmin}

The rate as a function of $\vmin$ behaves very differently for different $r/r_s$ as shown in Fig.~\ref{fig:vdf_rate_vmin}.  For low values of $r/r_s$, the change in detection rates between experiments can be much larger than the predictions of the SHM, e.g., the ratio between the rates of CoGeNT~\citep{2011PhRvL.106m1301A} to DAMA-I~\citep{2008EPJC...56..333B} is three times larger in our model than in the SHM.  This clearly motivates efforts to better constrain the scale radius of the Milky Way: comparing the scatter coming from measurements of VDF with the intrinsic physical differences among halos, the uncertainty on $r/r_s$ appears to be the dominant contribution to the uncertainty in event rates, especially for smaller $\vmin$.

\section{Discussion and Conclusion}
When deducing the direct detection event rate from cosmological simulations, the primary sources of uncertainty arise from: (i) finite particle sampling of the VDF, (ii) intrinsic scatter from physical processes that affect the VDF during the halo formation process (i.e.\ the halo-to-halo scatter), (iii) the quality of the fit and the validity of a smooth model, (iv) the observational constraint on $r/r_s$ for the Milky Way, (v) the variation of the VDFs in various directions at a fixed radius, and (vi) the impact of baryons. 
 
An important outcome of our analysis is that at present the scatter from (iv) is significantly larger than the corresponding scatter due to each of (i), (ii), and (iii), combined, by more than two orders of magnitude.  This is particularly important given that the observational constraint on the scale radius suggests the concentration $c=r_{\mathrm{vir}}/r_s$ is $10-20$~\citep{2002ApJ...573..597K,2012MNRAS.424L..44D}, which corresponds to $r_\odot/r_s\sim 0.15-0.6$~\citep{2008ApJ...684.1143X,2010ApJ...720L.108G,2010AJ....139...59B,2011ApJ...743...40B}.  Thus, although the distance from the Earth to the Galactic center is well known~\citep{2008ApJ...689.1044G,2009ApJ...692.1075G}, we find that the largest current theoretical uncertainty on the VDF is the uncertainty in $r/r_s$.

Our determination of the VDF represents an average over a spherical shell.  In reality, spherical asymmetry and substructures will affect the VDF and result in additional scatter along different directions.  In the {\sc Rhapsody} simulations, if we divide the spherical shell into several regions while maintaining enough particles (of the order 1000) in each analysis region, we find that this directional scatter is comparable to the halo-to-halo scatter, and that the combined scatter will be $10-40\%$ larger than only the halo-to-halo scatter, as illustrated in Fig.~\ref{fig:vdf_rate_vmin}.  Similar scatter is also seen in the {\sc Aquarius} Milky Way simulations~\citep{2009MNRAS.395..797V}.  This directional scatter will grow at larger radii because it is a consequence of substructures, tidal effects, and streams~\citep{2003MNRAS.339..834H,2011MNRAS.413.1419V,2011MNRAS.415.2475M,2012JCAP...08..027P}.  At present, we have no robust way to relate this scatter to direct observables, and in practice this directional scatter may be the most important uncertainty in determining the direct detection rates once other sources have been minimized. 

We have not yet investigated the impact of baryons; we expect that adiabatic contraction of dark matter halos would raise the velocity but preserve the shape of the VDF, so that our model will serve as a useful tool for these studies in the future.  Baryonic effects in isolated halos have been studied in the context of dark matter detection~\citep{2009ApJ...696..920B,2010JCAP...02..012L}; however, simulating a statistical sample of halos similar to what we consider here, with both sufficient resolution and realistic baryonic physics is not yet tractable.

The results presented here highlight the need to significantly improve the determination of the Milky Way scale radius. Although the concentration is now only weakly constrained with present data~\citep{2011ApJ...743...40B,2012MNRAS.424L..44D}, improvements will be forthcoming with spectroscopy and astrometry from large scale surveys~\citep{2012MNRAS.420.2562A}. Analysis along these lines will usher in a new era of complementarity between astronomical surveys and particle dark matter constraints deduced from terrestrial experiments.

\acknowledgments
  This work was supported by the U.S.~Department of Energy under
  contract number DE-AC02-76SF00515 and by a KIPAC Enterprise Grant.
  Mao is supported by a Weiland Family Stanford Graduate Fellowship. 
  We thank the Aspen Center for Physics and NSF Grant 1066293
  for hospitality during the workshop ``A Theoretical and
  Experimental Vision for Direct and Indirect Dark Matter
  Detection''.  We thank Anatoly Klypin and
  Joel Primack for providing access to the {\sc Bolshoi} simulations,
  and Peter Behroozi for the halo catalogs for both simulations.  We
  thank Tom Abel, Peter Behroozi, Michael Busha, Eduardo Rozo, 
  and Li-Cheng Tsai for useful discussion.  
  The {\sc Rhapsody} simulations were run using
  computational resources at SLAC; we gratefully acknowledge the
  support of Stuart Marshall, Amedeo Perazzo, and the SLAC
  computational team.

\end{document}